\documentclass[aps,prb,preprint]{revtex4}
\usepackage{CJK}
\usepackage{dcolumn}
\usepackage{amsmath}
\usepackage{amssymb}
\usepackage[dvips]{epsfig}  

\begin{document}
\begin{CJK*}{GBK}{}
\title{\bf  Realization of high-capacity hydrogen storage using carbon atomic chains: the role of terminations}
\author{Chun-Sheng Liu, Hui An, and Zhi Zeng\footnote{Corresponding author; zzeng@theory.issp.ac.cn.}}
\affiliation {Key Laboratory of Materials Physics, Institute of
Solid State Physics, Chinese Academy of Sciences, Hefei 230031, P.
R. China and \\Graduate School of the Chinese Academy of Sciences}

\date{\today}

\begin{abstract}
The capacity of carbon atomic chains with different terminations for
hydrogen storage is studied using first-principles density
functional theory calculations. Unlike the physisorption of H$_2$ on
the H-terminated chain, we show that two Li (Na) atoms each capping
one end of the carbon chain can hold ten H$_2$ molecules with
optimal binding energies for room temperature storage. The
hybridization of the Li 2\emph{p} states with the H$_2$ $\sigma$
orbitals contributes to the H$_2$ adsorption. However, the binding
mechanism of the H$_2$ molecules on Na arises only from the
polarization interaction between the charged Na atom and the H$_2$.
Moreover, additional H$_2$ molecules can be bound to the carbon
atoms at the chain ends due to the charge transfer between Li
2\emph{s}2\emph{p} (Na 3\emph{s}) and C 2\emph{p} states.
Importantly, dimerization of these isolated metal-capped chains does
not affect the hydrogen binding energy significantly. In addition, a
single chain can be stabilized effectively by termination on the
C$_{60}$ clusters. With a hydrogen uptake of $>$ 10 wt \% on
Li-coated C$_{60}$-C$_n$-C$_{60}$ (\emph{n} = 5, 8), the
Li$_{12}$C$_{60}$-C$_n$-Li$_{12}$C$_{60}$ complex, without reducing
the number of adsorbed H$_2$ molecules per Li, can serve as better
building blocks of polymers than the (Li$_{12}$C$_{60}$)$_2$ dimer.
These findings suggest a new route to design cluster-assembled
storage materials based on terminated \emph{sp} carbon chains.
\end{abstract}


\maketitle


\section{\bf Introduction}
The development of economical hydrogen energy critically depends on
finding efficient and safe media that can store hydrogen with high
gravimetric and volumetric density.\cite{ref01} The storage
materials should also reversibly adsorb/desorb H$_2$ under ambient
thermodynamic conditions. These criteria not only limit the choice
of storage materials to those composed of light elements, but also
require the hydrogen adsorption energy to lie between the
physisorbed and chemisorbed states (0.1$\sim$0.2 eV) \cite{ref02}.

Carbon nanostructures with \emph{sp}$^2$-like bonding
\cite{ref03,ref04,ref05,ref06,ref07,ref08} have long been expected
to be promising storage materials due to their light weights and
large surface areas. However, hydrogen molecules bind weakly to
pristine carbon materials via van der Waals interactions
\cite{ref03}. A new approach, coating the carbon surfaces with early
transition metals (TMs), has recently been shown to improve storage
performance \cite{ref04,ref05}. These complexes not only enhance the
interaction between hydrogen molecules and TMs through the Kubas
interaction (hybridization of the \emph{d} states with the H$_2$
states) \cite{ref09}, but also meet the U.S. Department of Energy
goal of obtaining a gravimetric capacity of 9 wt \% by the year
2015. Unfortunately, preferable clustering of TM atoms on the carbon
nanomaterial surfaces results in poor reversibility \cite{ref10}.
Hence, the adsorption energy of metal atoms to the substrate should
be larger than the cohesive energy of the bulk metal. The use of
alkali metals (AMs) and alkaline-earth metals such as Li
\cite{ref11} and Ca \cite{ref12} to produce uniform coating has been
proposed, as their cohesive energies are substantially smaller than
those of TMs ($\sim$4 eV), but the binding energy of H$_2$ molecules
in the Li$_{12}$C$_{60}$ system is too small (0.075 eV/H$_2$) for
room temperature applications \cite{ref11}.

In the search for a more feasible high-capacity storage medium, we
now turn to the \emph{sp}-hybridized carbon chains, which are
considered as the important building blocks in nano-electronics and
nano-mechanics. The reasons for choosing carbon chains in our study
are the following: Due to their high reactivity, bare carbon chains
can be stabilized by termination on a wide variety of atoms and
groups \cite{ref13}. Isolated \emph{sp} chains terminated with
hydrogen atoms, i.e., polyynes (with alternating singlet-triplet
bonds) and cumulenes (with double C-C bonds), have already been
synthesized in their basic forms \cite{ref14}. In addition, chains
with metal atoms connected to the ends have been generated
\cite{ref15}, and their magnetic, electronic, and transport
properties have been studied extensively \cite{ref16,ref17}. In
particular, the hybrid \emph{sp}+\emph{sp}$^2$ carbon-based systems
\cite{ref18}(with linear chains connecting \emph{sp}$^2$-type
fragments), representing typical interfaces in realistic
nanostructures, exhibit exceptional physical and chemical
properties. Recent experiments have realized single carbon chains
bridging graphene nanoribbons \cite{ref19}, and produced junctions
between a single carbon chain and two single-wall carbon nanotubes
(SWCN)\cite{ref20}.

In this study, we conduct theoretical studies of hydrogen storage
media consisting of H, AMs, and C$_{60}$ terminated cumulenes or
polyynes. We show that hydrogen molecules prefer to be physisorbed
rather than chemisorbed on the H-terminated carbon chains. Thus Li
(Na) atoms capped carbon chains are chosen to improve the hydrogen
storage performance. As a result, H$_2$ molecules are adsorbed not
only on the Li or Na atoms, but also on the carbon chains. In
addition, a linear periodic polymer, composed of Li atoms and five
(eight)-atom carbon chain bridging C$_{60}$ fullerenes, can store up
to $\sim$10 wt \% of H$_2$ molecules with an average binding energy
of 0.15 (0.14) eV/H$_2$. Our results not only advance our
understanding of H$_2$ adsorption on terminated carbon chains, but
also offer a new avenue for efficient hydrogen storage.

\section{\bf Computational details}
Numerical calculations were performed using spin-polarized density
functional theory (DFT) with the Perdew-Wang (1991)
exchange-correlation function \cite{ref21}, as implemented in the
DMol$^3$ package (Accelrys Inc.). \cite{ref22} A double numerical
atomic orbital augmented by \emph{d}-polarization functions (DNP)
was employed as the basis set. In the self-consistent field
calculations, the electronic-density convergence threshold was set
to 1$\times$10$^{-6}$ electron/\AA$^3$. Geometric optimization was
performed with convergence thresholds of 10$^{-5}$ Ha for the
energy, 2$\times$10$^{-3}$ Ha/\AA $ $ for the force, and 10$^{-4}$
\AA $ $ for the atomic displacements. We performed normal-mode
analysis of the obtained structures to ensure that the structures
optimized without any symmetry constraints were true minima of the
potential-energy surface. To validate the above convergence
parameters, the bond length and vibrational frequency of the C$_2$
dimer were calculated to be 1.312 \AA$ $ and 1647 cm$^{-1}$,
respectively, which agree well with the corresponding experimental
values of 1.31 \AA $ $ and 1641 cm$^{-1}$ \cite{ref23}. The
calculated bond length of the H$_2$ molecule (0.748 \AA) is very
close to the experimental value of 0.741 \AA$ $.

\section{Results and discussion}

\subsection{H$_2$ adsorption on H-terminated carbon chains}

We first consider the physisorption of a hydrogen molecule on the
standard cumulene (C$_5$H$_4$) and polyyne (C$_8$H$_2$). In Fig. 1,
the lowest energy configuration of a H$_2$ molecule is located above
C-C bond with the H-H molecular axis between parallel and
perpendicular to the C-C bond. The adsorption energies of H$_2$ to
C$_5$H$_4$ and C$_8$H$_2$ are about 0.05 eV and 0.08 eV,
respectively, resulting from van der Walls interactions. We also
calculated the adsorption energies of H$_2$ to
C$_{2\emph{n}+1}$H$_4$ (\emph{n} = 3, 4) and C$_{2\emph{n}}$H$_2$
(\emph{n} = 2, 3, 5), and found that the physisorption energies did
not vary significantly with the chain lengths.

Next, we investigate the H$_2$ dissociated chemisorption on
C$_5$H$_4$ or C$_8$H$_2$ by optimizing the structure with two H
atoms placed close to two adjacent C atoms (i.e., C1 and C2). Upon
the atomic hydrogen adsorption, transformation of the \emph{sp}
hybridization of C1-C2 bond to the \emph{sp}$^2$ hybridization
results in a change of the bond lengths and angles. A significant
elongation of the C1-C2 bond length in the chemisorption state
indicates that this bond is weakened.

We apply the nudged elastic band (NEB) method to study the
minimum-energy path (MEP) \cite{ref24} of the hydrogen molecule
dissociation on the carbon chains. The image number considered is 16
to ensure that the obtained MEP is correct. The physisorption of the
hydrogen molecule is chosen as the initial state (PS) and the
chemisorption of two hydrogen atoms as the final state (CS), as
depicted in Fig. 1. From PS to CS, high barriers of 3.81 eV and 4.32
eV must be overcome. This indicates that the dissociation process of
H$_2$ on C$_5$H$_4$ or C$_8$H$_2$ is very difficult in kinetics.

\subsection{H$_2$ adsorption on Li (Na)-capped carbon chains}

The above results have demonstrated that the interaction between
pristine carbon chains and the hydrogen molecule is very weak, such
storage systems may have to be operated at below ambient
temperature. Now the question arises: can the storage performance of
carbon chains be improved by using alternative capping elements? We
then consider several kinds of short atomic chains (C$_{\emph{n} =
5-10}$) terminated with two alkali metal atoms. Our full structural
relaxations reveal that the linear Li$_2$C$_n$ and Na$_2$C$_n$
structures are more stable and energetically favorable relative to
the zigzag structures. Figure 2a illustrates that the average
binding energies per Li (Na) atom with C$_n$ chains show an even-odd
oscillatory behavior. These binding energies are obviously much
larger than those of Li (1.63/1.52 eV) and Na (1.11/1.41 eV) on
graphene/fullerene \cite{ref25,ref26}. In addition, the
configurations of Li$_2$C$_n$ and Na$_2$C$_n$ complexes exhibit
significant differences for odd or even values of \emph{n}. For
instance, when \emph{n} = 5, the C-C distances, as depicted in Figs
.2b and 2d, are rather uniform. In contrast, the bond lengths (Figs.
2c and 2e) alternate for \emph{n} = 8. This behavior implies that
the bonding patterns of Li(Na)$_2$C$_{2\emph{n}+1}$ and
Li(Na)$_2$C$_{2\emph{n}}$ are the same as those of C$_{2\emph{n}+1}$
and C$_{2\emph{n}}$, respectively.

Note that the induced positive charge of the Li atom is smaller than
that of Na, which is not compatible with the order of their binding
energies (Fig. 2a). For example, each Li (Na) atom carries 0.39
(0.62) \emph{e} and 0.43 (0.66) \emph{e} in Li$_2$C$_5$
(Na$_2$C$_5$) and Li$_2$C$_8$ (Na$_2$C$_8$), respectively. We find
that Li 2\emph{p} orbitals become partly filled due to the carbon
chain back-donating some electrons to Li, whereas Na still has empty
3\emph{p} orbitals. This discrepancy means that the strong bonding
between the Li atom and carbon chains is not simply ionic,
distinctly different from the ionic bonding as the Na atom is
attached to the chains. Figures 2f and 2g clearly show that the
2\emph{p} state of Li and the C \emph{p} orbitals are hybridized for
the binding of Li on the C$_5$ and C$_8$ chain. Since the
\emph{p}-\emph{p} hybridization is stronger than the
\emph{s}-\emph{p} hybridization, the binding energies of Li atoms to
the carbon chains are larger than those of Na atoms in the
Na$_2$C$_n$ complexes. This bonding mechanism has also been observed
in Li coated boron-doped graphenes \cite{ref27}. In detail, the
hybridization of the Li \emph{p}$_y$ orbital with the lowest
unoccupied molecular orbitals (LUMO) of C$_5$ gives rise to
$\pi$-bonds (Fig. 2f), whereas the overlap between Li \emph{p}$_x$
and C \emph{p}$_x$ orbitals in Li$_2$C$_8$ forms $\sigma$-bonds
(Fig. 2g). Because the $\pi$ bond is generally weaker than the
$\sigma$ bond, it is understandable that Li atoms bind more strongly
to C$_{2\emph{n}}$ than that to C$_{2\emph{n}+1}$ (as revealed in
Fig. 2a).

We now turn to the interaction between the metal-capped carbon
chains and hydrogen. The results of our calculations present that
one Li (Na) atom can attach 5 H$_2$ molecules in the Li$_2$C$_5$
(Na$_2$C$_5$) and Li$_2$C$_8$ (Na$_2$C$_8$) complexes, suggesting
that the maximum number of adsorbed H$_2$ molecules does not depend
on the chain types. The optimized geometries for 5 H$_2$ molecules
adsorbed on Li$_2$C$_n$ and Na$_2$C$_n$ (\emph{n} = 5, 8), as
illustrated in Figs. 3a-3d, demonstrate that the H-H distances are
slightly enlarged compared with the free H$_2$ bond length (0.74
\AA). All the side H$_2$ molecules tend to tilt toward the chain so
that one of two H atoms of adsorbed H$_2$ molecules becomes
relatively closer to the metal atom. To understand the binding
mechanism of H$_2$ on Li- or Na-capped carbon chains, we employ the
Mulliken charge analysis. In detail, charge transfer from the
$\sigma$ orbital of a side H$_2$ to the attached Li atom ($\sim$0.08
\emph{e}) is four times larger than that to the Na atom ($\sim$0.02
\emph{e}). Therefore the binding of H$_2$ to the Na atom arises
basically from the induced polarization of the H$_2$ molecules by
the Na atom. However the polarization interaction is not the only
factor responsible for H$_2$ binding to Li due to the average
binding energies of H$_2$ molecules (Figs. 3e and 3f) in
Li$_2$C$_5$(H$_2$)$_5$ (0.19 eV/H$_2$) and Li$_2$C$_8$(H$_2$)$_5$
(0.19 eV/H$_2$) being slightly larger than those in
Na$_2$C$_5$(H$_2$)$_5$ (0.15 eV/H$_2$) and Na$_2$C$_8$(H$_2$)$_5$
(0.14 eV/H$_2$). The analysis of molecular-orbital coefficients for
the Li$_2$C$_5$(H$_2$)$_5$ and Li$_2$C$_8$(H$_2$)$_5$ complexes
shows that the hybridization of the Li 2\emph{p} and 2\emph{s}
states with the H$_2$ $\sigma$ orbitals (see Figs. 3g and 3h) also
contribute to the H$_2$ adsorption. The consequences of the orbital
interactions are as well reflected in the charge variations of the
Li. The effective charge of the Li atom in both Li$_2$C$_5$ and
Li$_2$C$_8$ complexes varies from 0.32 to 0.06 \emph{e} as the
number of hydrogen molecules increases from 1 to 5. As a result, the
binding energies of H$_2$ molecules to the Li atom slightly decrease
with the increasing number of H$_2$ molecules (see Figs. 3e and 3f).

It is interesting to note that the H$_2$ molecules are adsorbed not
only on the Li or Na atoms, but also on the carbon chains. Figure 4
shows that additional 8 H$_2$ molecules can be attached to
Li$_2$C$_5$(H$_2$)$_{10}$, Li$_2$C$_8$(H$_2$)$_{10}$,
Na$_2$C$_5$(H$_2$)$_{10}$, and Na$_2$C$_8$(H$_2$)$_{10}$ complexes,
resulting in hydrogen capacities of 32.7, 24.7, 25.4, and 20.2 wt
\%, respectively. The average binding energies of these 8 H$_2$
molecules to Li$_2$C$_n$(H$_2$)$_{10}$ and Na$_2$C$_n$(H$_2$)$_{10}$
(\emph{n} = 5, 8) are 0.11 and 0.12 eV/H$_2$, respectively, which
lie within the required range of 0.1-0.2 eV. The essential reason
for hydrogen attachment to the chain is that carbon atoms
participate in the electron transfer, and thus the carbon chain
differs from the pristine one. For instance, a C atom at the end of
Li$_2$C$_5$ (Na$_2$C$_5$) obtains the largest amount of electrons
from Li (Na) and carries -0.37 \emph{e} (-0.46 \emph{e}). Very
importantly, the charge on the carbon atoms does not vary during
hydrogen molecules adsorbed on metal atoms. Then the electric field
around the C atoms, generated by the substantial charge
redistribution, can polarize the hydrogen molecules. This is
confirmed through the plot of the deformation electron density (Fig.
4) which exhibits the localized characteristics of H-H bonds and no
charge transfer between H and C atoms, different from the charge
distribution of the side H$_2$ molecules bound to Li atoms. The H-H
electric dipole vector should be parallel to the electric field
direction, thus the H$_2$ molecules align vertical to the chain in
such electric field.

The above results are just promising for the isolated metal-capped
carbon chains. We therefore put our further considerations on the
cluster-assembled materials by using these metal-capped carbon
chains as building blocks. The geometry optimization of
M$_2$C$_{\emph{n} = 5, 8}$ (M = Li, Na) dimers was performed by
starting with several initial configurations. As shown in Fig. 5,
two M$_2$C$_\emph{n}$ complexes have formed stable dimers through
C-C bonds or M-C bonds but not M-M bonds due to the repulsion
between the like charges. The (Li$_2$C$_5$)$_2$ dimer has a binding
energy of 3.75 eV which is substantially smaller than 7.70 eV
corresponding to the (Li$_2$C$_8$)$_2$ dimer due to C-C bonds
formation between two Li$_2$C$_8$ complexes. On the contrary, the
binding energy of the (Na$_2$C$_5$)$_2$ dimer (2.72 eV) is larger
than that of the (Na$_2$C$_8$)$_2$ dimer (1.74 eV). However the
dimerization affects the number of adsorbed hydrogen molecules, as
smaller charge on the metal atom reduces the polarization effect of
the metal cation on the hydrogen molecules. As a result, the
gravimetric densities obtained by the (Li$_2$C$_5$)$_2$,
(Li$_2$C$_8$)$_2$, (Na$_2$C$_5$)$_2$, and (Na$_2$C$_8$)$_2$ dimers
are lowered to 14.0, 9.8, 13.1, and 11.3 wt \%, respectively.
Importantly, dimerizations do not change the binding energies of the
H$_2$ on metal atoms significantly, and therefore should not affect
the desorption temperature.

\subsection{H$_2$ adsorption on Li$_{12}$C$_{60}$-C$_\emph{n}$-
Li$_{12}$C$_{60}$}

From above, we have seen that the type of chain-termination turns
out to influence the hydrogen adsorption performance. Here we have
considered the \emph{sp}$^2$-terminated C$_\emph{n}$ chains as
potential storage media by choosing C$_{60}$ fullerene as the
end-capping candidate, because it not only mimics the inner SWNT
caps \cite{ref20} but also can hold metal atoms to attach hydrogen
molecules. We find that C$_\emph{n}$ chains, regardless of \emph{n}
being odd or even, energetically prefer to bind the common bond of
two hexagons ([6,6]), resulting in the
C$_{60}$-C$_{2\emph{n}+1}$-C$_{60}$ and
C$_{60}$-C$_{2\emph{n}}$-C$_{60}$ complexes with \emph{D}$_{2d}$ and
\emph{D}$_{2h}$ symmetry, respectively. As show in Figs. 6a and 6b,
the bond lengths of the C$_5$ (C$_8$) chain in
C$_{60}$-C$_5$-C$_{60}$ (C$_{60}$-C$_8$-C$_{60}$) imply a polyyne
(cumulene) structure. The binding energy is calculated by
subtracting the equilibrium total energy of the
C$_{60}$-C$_\emph{n}$-C$_{60}$ complex from the sum of the total
energy of isolated chain (C$_n$) and of C$_{60}$ clusters. The
structure of two C$_{60}$ clusters bridged by the C$_5$ (C$_8$)
chain with a binding energy of 4.46 eV (4.94 eV) is more stable than
the dumbbell-like structure of C$_{60}$ dimers \cite{ref28}. The
stability of C$_{60}$-C$_\emph{n}$-C$_{60}$ (\emph{n} = 5, 8) has
been further verified by carrying out \emph{ab initio} molecular
dynamics simulations at 800 K with a time step of 1 fs. After
running 5000 steps, the geometry is still kept although the chain is
slightly distorted, suggesting that the
C$_{60}$-C$_\emph{n}$-C$_{60}$ complexes are stable. This is
confirmed by recent theoretical results that C$_{60}$-C$_8$-C$_{60}$
can be stable up to a temperature of 1000 K \cite{ref20}.

Considering the Li atoms interacting with the
C$_{60}$-C$_\emph{n}$-C$_{60}$ (\emph{n} = 5, 8) complex, it is
found that the pentagonal face of C$_{60}$ is the preferred site. A
close examination of the geometry of the
Li$_{12}$C$_{60}$-C$_\emph{n}$-Li$_{12}$C$_{60}$ configuration, as
depicted in Figs. 6c and 6d, illustrates that the 12 Li atoms
adsorbed on each C$_{60}$ cluster are divided into two sets. The
Mulliken charge analysis presents that the Li 2\emph{p} orbitals are
partly filled. Thus the hybridization between the Li 2\emph{p} and C
2\emph{p} orbitals strengthens the Li binding to C$_{60}$. The
average binding energy per Li atom with C$_{60}$ in
Li$_{12}$C$_{60}$-C$_5$-Li$_{12}$C$_{60}$ (1.74 eV) or
Li$_{12}$C$_{60}$-C$_8$-Li$_{12}$C$_{60}$ (1.76 eV) is larger than
that in the isolated Li$_{12}$C$_{60}$ (1.71 eV) and the Li bulk
cohesive energy (1.63 eV experimental value). Hence the clustering
of Li atoms will not occur in our models.

We now enter the next phase of our calculation, namely, to discuss
the possible number of hydrogen molecules that can be bound to the
Li$_{12}$C$_{60}$-C$_5$-Li$_{12}$C$_{60}$ and
Li$_{12}$C$_{60}$-C$_8$-Li$_{12}$C$_{60}$ structures. Previous
studies have suggested that 5 H$_2$ molecules can be bound to each
Li atom in the isolated Li$_{12}$C$_{60}$ cluster, whereas the
binding energy of the fifth H$_2$ molecule is only 0.06 eV
\cite{ref12}. Therefore we consider if each Li atom can adsorb 4
H$_2$ molecules in both Li$_{12}$C$_{60}$-C$_5$-Li$_{12}$C$_{60}$
and Li$_{12}$C$_{60}$-C$_8$-Li$_{12}$C$_{60}$ structures,
corresponding to 10.3 and 10.1 wt \% gravimetric densities,
respectively. The average binding energies of 96 H$_2$ molecules
with Li$_{12}$C$_{60}$-C$_5$-Li$_{12}$C$_{60}$ (Fig. 7a) and
Li$_{12}$C$_{60}$-C$_8$-Li$_{12}$C$_{60}$ (Fig. 7a) are 0.16 eV. The
equilibrium H-H bond lengths are 0.755 \AA, which is comparable to
that in Li$_{12}$C$_{60}$(H$_2$)$_{60}$, namely, 0.753 \AA. The
shortest and longest distances between the Li atom and H atoms are
observed to be of the order of 2.330-2.750 \AA.

As mentioned above, a single carbon chain bridging two
Li$_{12}$C$_{60}$ clusters does not change the number of adsorbed
hydrogen molecules per Li. In contrast, a (Li$_{12}$C$_{60}$)$_2$
dimer would lower the hydrogen storing capacity because some of the
Li atoms linking Li$_{12}$C$_{60}$ clusters may not be able to
attach hydrogen molecules \cite{ref12}. Now readers have to ask
whether the Li$_{12}$C$_{60}$-C$_\emph{n}$-Li$_{12}$C$_{60}$
structures can form the building blocks of a new kind of solid
storage materials? To address this question, we first consider the
adsorption of Li atoms on linear periodic
C$_{60}$-C$_\emph{n}$-C$_{60}$ polymers. After full symmetry
unrestricted geometry optimization,
Li$_{12}$C$_{60}$-C$_\emph{n}$-Li$_{12}$C$_{60}$ units can maintain
their structural identity. The binding energy of Li atoms to a
linear periodic C$_{60}$-C$_5$-C$_{60}$ (C$_{60}$-C$_8$-C$_{60}$)
polymer is 1.80 (1.76) eV/Li, respectively, indicating that the
polymerization has not affected the binding strength of Li to
C$_{60}$. On the experimental side, one-dimensional \cite{ref28} or
two-dimensional \cite{ref29} C$_{60}$ polymers, the hexagonal faces
interact with each other, have been synthesized by several methods.
For the fabrication of C$_{60}$-C$_\emph{n}$-C$_{60}$ polymer, we
propose a possible route to use the SWNT/single-carbon-chain
molecular junctions \cite{ref20} as the initial materials. Second,
each Li atom in the polymer composed of
Li$_{12}$C$_{60}$-C$_5$-Li$_{12}$C$_{60}$
(Li$_{12}$C$_{60}$-C$_8$-Li$_{12}$C$_{60}$), shown in Fig. 8, can
hold 4 H$_2$ molecules with an average binding energy of 0.15
eV/H$_2$ (0.14 eV/H$_2$), resulting in the maximum gravimetric
density of 10 wt \% (9.6 wt \%). This not only meets the gravimetric
density target set by the U.S. Department of Energy for the year
2015, but the hydrogen binding energy is also optimal for the system
operation under ambient thermodynamic conditions.

\section{\bf Conclusions}

In conclusion, the adsorption of H$_2$ molecules on H, AMs, and
C$_{60}$ terminated carbon chains was studied using all-electron DFT
calculations. It is energetically favorable for H$_2$ to physisorb
on the H terminated chains. However, for Li or Na atoms capping the
chain ends, H$_2$ molecules are adsorbed not only on the metal atoms
but also on the carbon atoms with the required adsorption energy.
Although the dimerization of M$_2$C$_\emph{n}$ (M = Li, Na) lower
the hydrogen uptake, the binding energy of the H$_2$ on metal atoms
is not affected significantly. Thus one may assemble these
metal-capped carbon chains to synthesize efficient storage media or
catalysts. Furthermore, we propose a linear periodic polymer,
consisting of Li$_{12}$C$_{60}$-C$_5$-Li$_{12}$C$_{60}$
(Li$_{12}$C$_{60}$-C$_8$-Li$_{12}$C$_{60}$), can store up to
$\sim$10 wt \% of H$_2$ molecules with an average binding energy of
0.15 (0.14) eV/H$_2$. We believe that current results presented here
can be applied to other carbon chain bridging metal-coated
\emph{sp}$^2$ materials. We also hope these results may provide a
useful reference for the design of bulk hydrogen storage materials
in the laboratory.

\section{\bf Acknowledgments}

This work was supported by the National Science Foundation of China
under Grant no. 10774148, the special funds for the Major State
Basic Research Project of China (973) under Grant no. 2007CB925004,
and the Knowledge Innovation Program of the Chinese Academy of
Sciences. Part of the calculations were performed at the Center for
Computational Science of CASHIPS. We thank P. Jena for useful
suggestions.

\clearpage
\newpage
\newpage


\begin{figure}[htbp]

\includegraphics[width=0.5\textwidth]{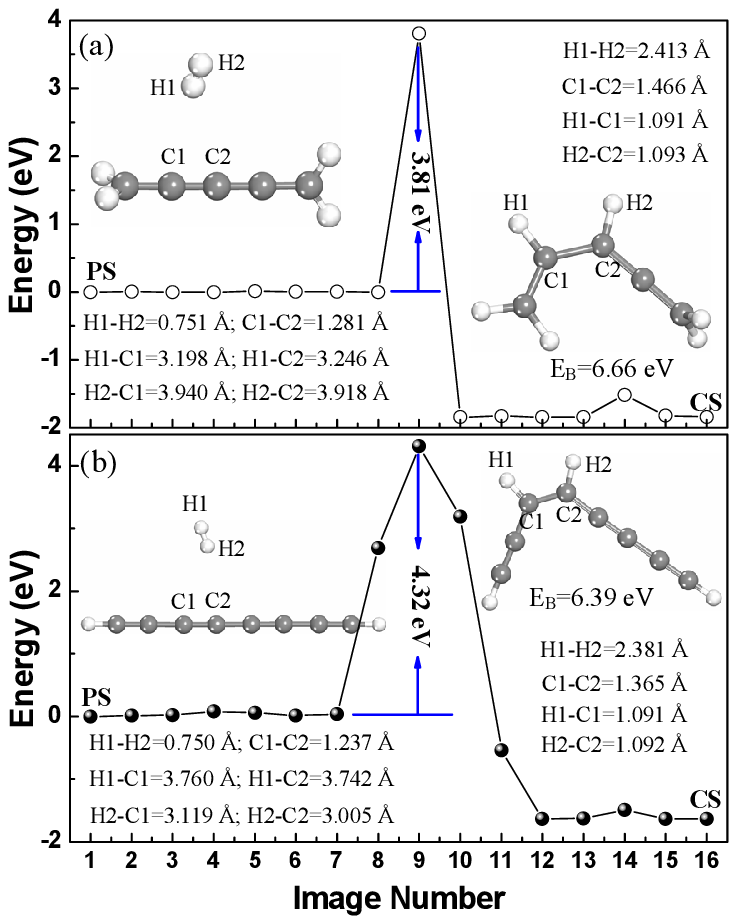}
\caption{The MEPs for the dissociation process of H$_2$ on the (a)
C$_5$H$_4$ and (b) C$_8$H$_2$. The energy of the initial structure
(PS) is set as zero. The relevant bond distances and binding energy
(E$_B$) are also given.}
\end{figure}
\clearpage
\newpage

\begin{figure}[!ht]

\includegraphics[width=0.5\textwidth]{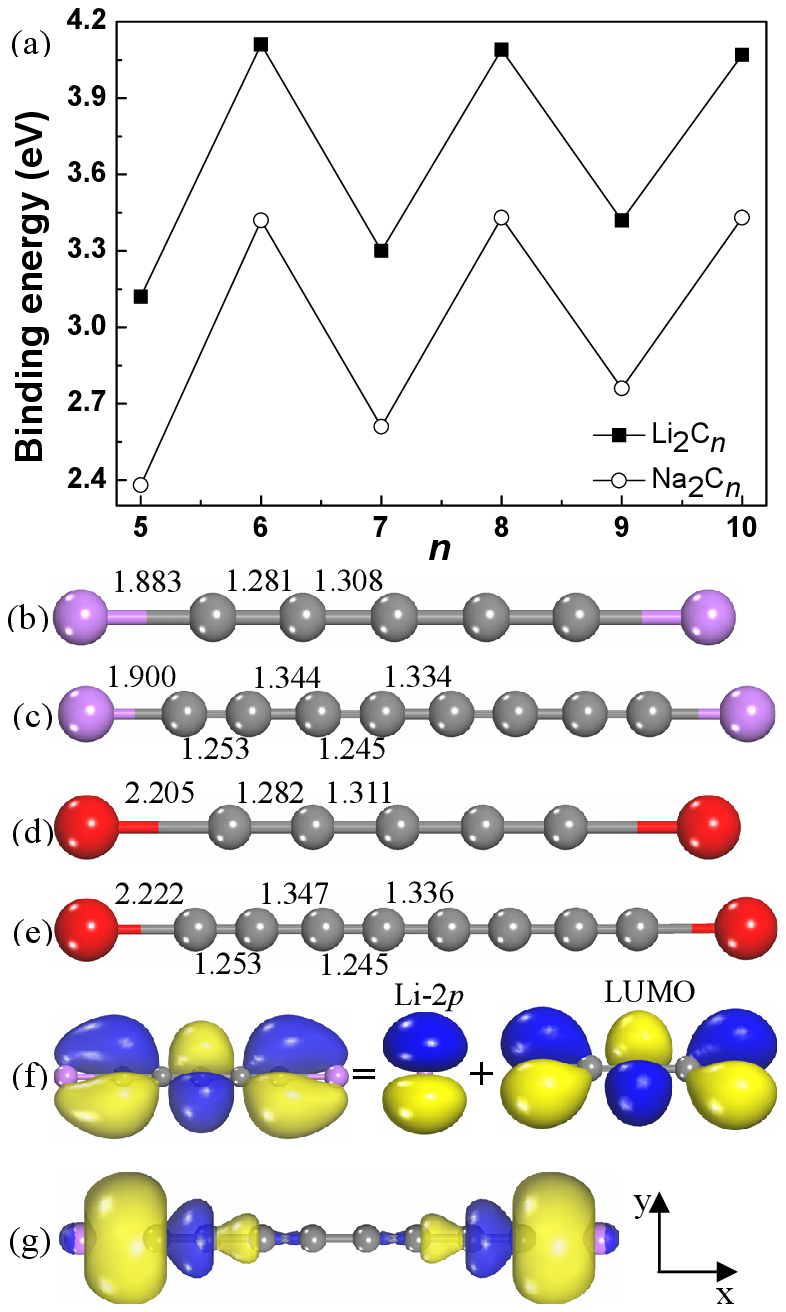}
\caption{(a) Average binding energies of two Li or Na atoms on
C$_\emph{n}$ as a function of the number of carbon atoms. The
optimized bond lengths (in \AA) of (b) Li$_2$C$_5$, (c) Li$_2$C$_8$,
(d) Na$_2$C$_5$, and (e) Na$_2$C$_8$. The large (red), medium
(purple), and small (grey) balls represent the Na, Li, and C atoms,
respectively. The panel (f) shows the Li$_2$C$_5$ bonding orbital
results from the hybridization of the Li 2\emph{p} orbital and C$_5$
LUMO. (g) The bonding orbital for the Li atoms attached to C$_8$.
The isovalue equals 0.015 e/\AA$^3$. The two colors denote $\pm$
signs of the wave function. }

\end{figure}
\clearpage
\newpage
\clearpage
\newpage

\begin{figure}[!ht]

\includegraphics[width=0.8\textwidth]{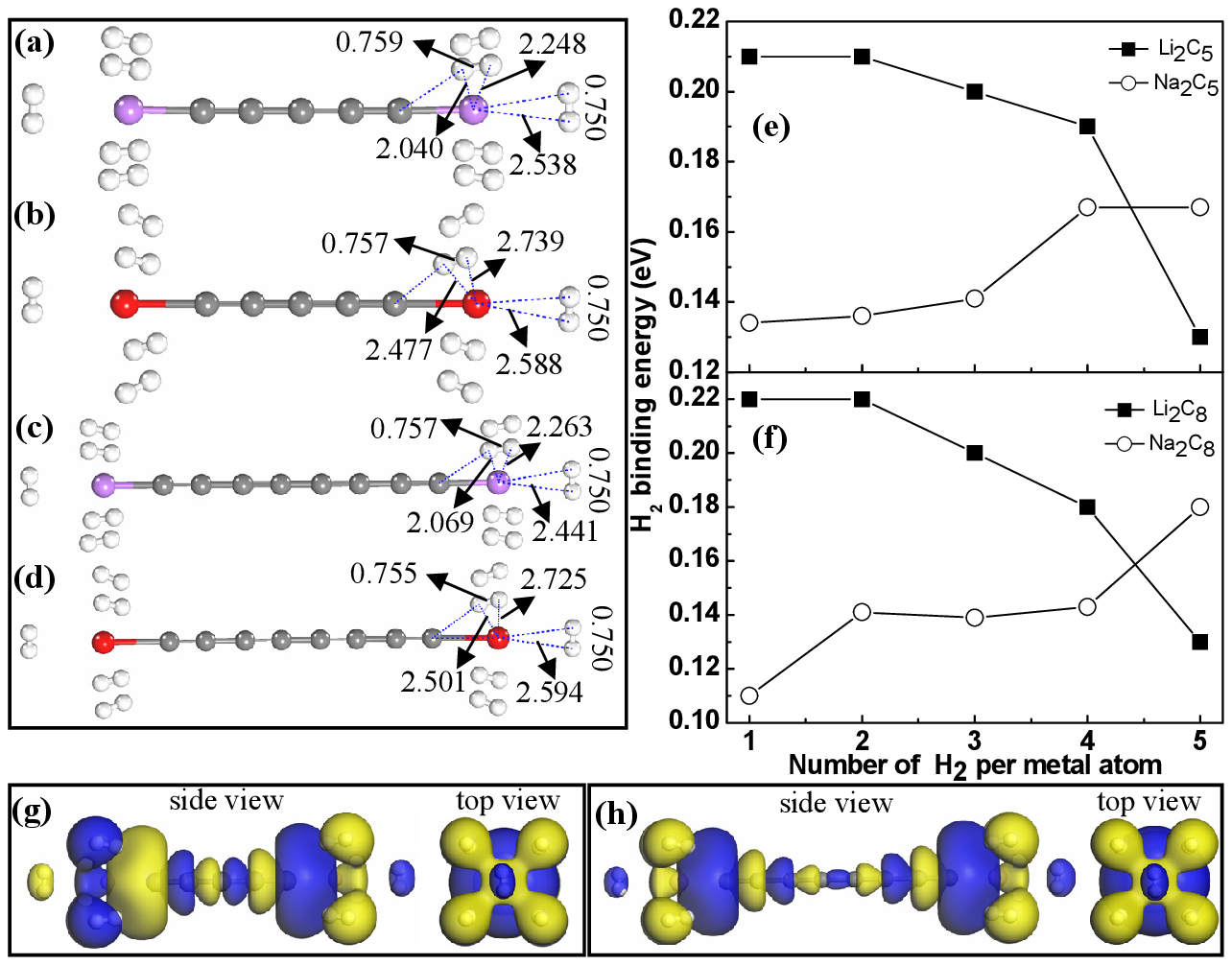}
\caption{Optimized configurations of H$_2$ molecules adsorbed on (a)
Li$_2$C$_5$, (b) Na$_2$C$_5$, (c) Li$_2$C$_8$, and (d) Na$_2$C$_8$
along with the typical bond lengths (in \AA). Panels (e) and (f)
show the hydrogen binding energies with successive additions of
H$_2$ molecules to each metal atom in M$_2$C$_\emph{n}$ (M = Li, Na;
\emph{n} = 5, 8). Panels (g) and (h) show the orbitals for the side
hydrogen molecules hybridized with the Li 2\emph{p} orbitals in
Li$_2$C$_5$(H$_2$)$_5$ and Li$_2$C$_8$(H$_2$)$_5$, respectively. The
isovalue equals 0.01 e/\AA$^3$.}

\end{figure}
\clearpage
\newpage
\clearpage
\newpage

\begin{figure}[!ht]

\includegraphics{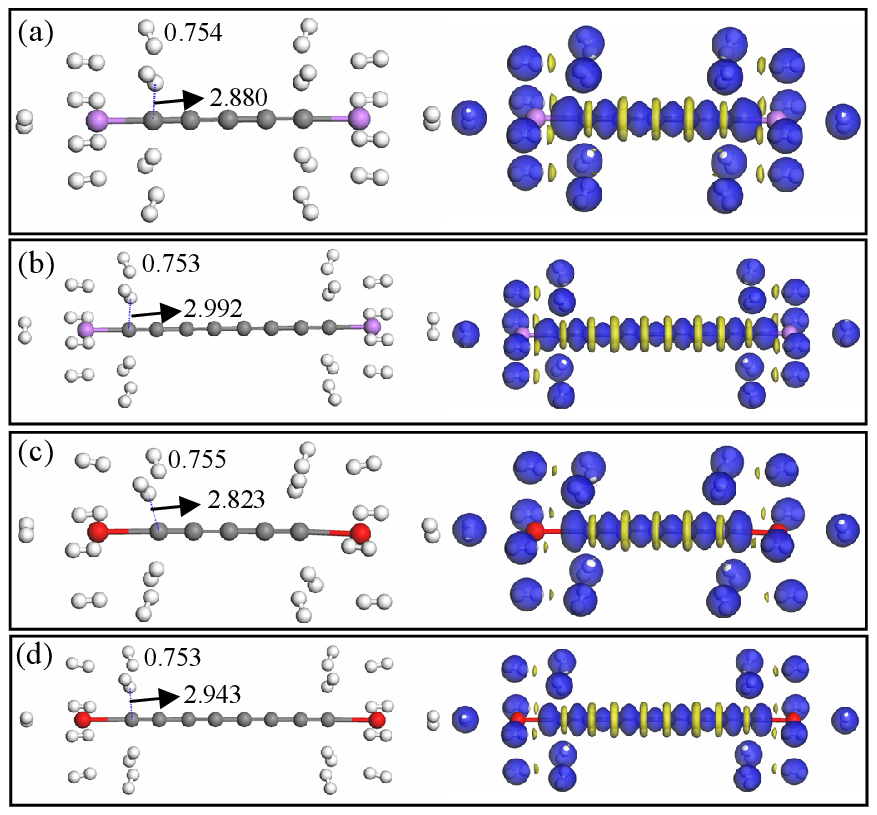}
\caption{Final optimized geometries of (a) Li$_2$C$_5$(H$_2$)$_5$,
(b) Li$_2$C$_8$(H$_2$)$_5$, (c) Na$_2$C$_5$(H$_2$)$_5$, and (d)
Na$_2$C$_8$(H$_2$)$_5$ holding additional eight H$_2$ molecules. The
relevant bond distances (in \AA) are given. The right panels (a),
(b), (c), and (d) show the corresponding deformation electron
densities (molecular charge densities minus atomic charge
densities). The deformed density marked in blue corresponds to the
region that contains excess electrons, while that marked in yellow
indicates electron loss. The isovalue equals 0.1 e/\AA$^3$.}

\end{figure}
\clearpage
\newpage

\begin{figure}[!ht]

\includegraphics{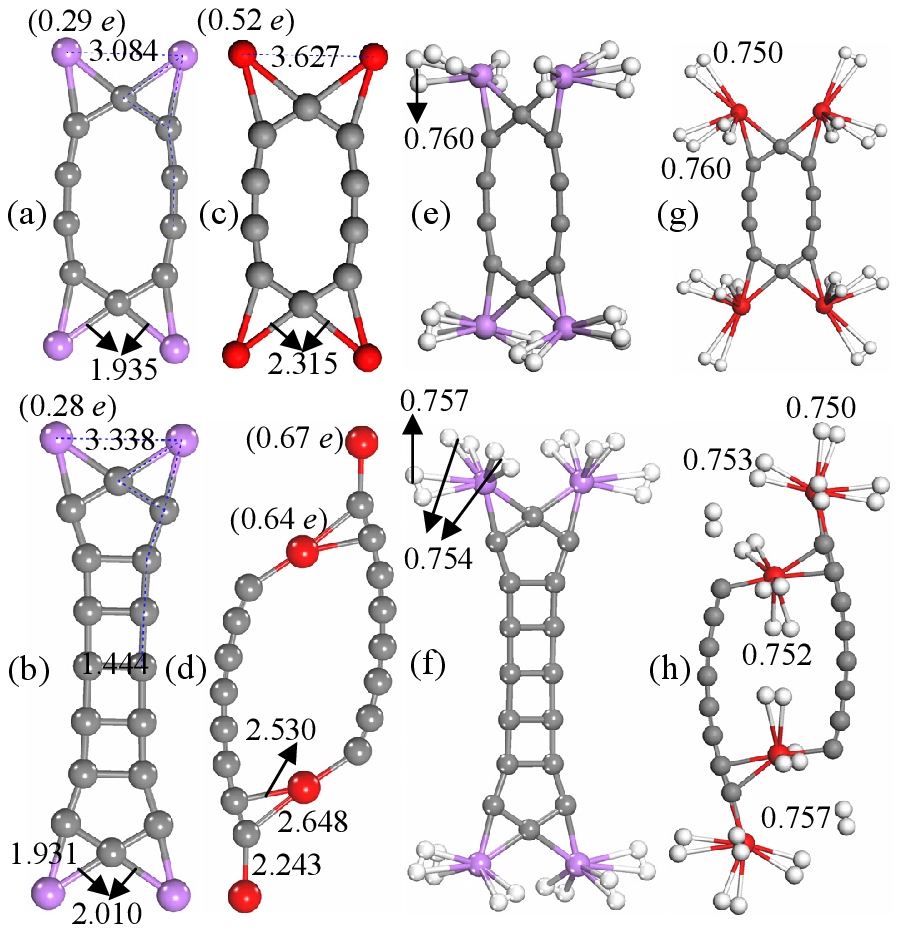}
\caption{Atomic configurations of dimers of (a) Li$_2$C$_5$, (b)
Li$_2$C$_8$, (c) Na$_2$C$_5$, and (d) Na$_2$C$_8$. Geometries of (e)
Li$_2$C$_5$, (f) Li$_2$C$_8$, (g) Na$_2$C$_5$, and (h) Na$_2$C$_8$
dimers respectively holding 12, 12, 12, and 18 H$_2$ molecules. The
numbers in parentheses refer to charges on metal atoms, while other
numbers refer to bond lengths in \AA.}

\end{figure}
\clearpage
\newpage

\begin{figure}[!ht]

\includegraphics[width=0.8\textwidth]{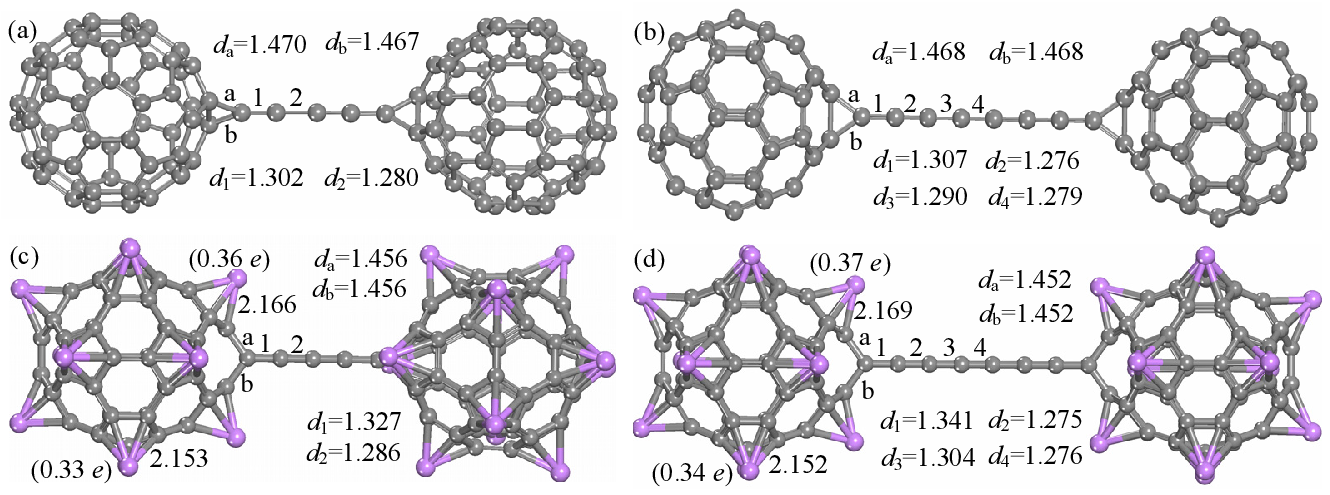}
\caption{Atomic configurations of (a) C$_{60}$-C$_5$-C$_{60}$, (b)
C$_{60}$-C$_8$-C$_{60}$, (c)
Li$_{12}$C$_{60}$-C$_5$-Li$_{12}$C$_{60}$, and (d)
Li$_{12}$C$_{60}$-C$_8$-Li$_{12}$C$_{60}$. The numbers in
parentheses denote charges on the Li atoms. The typical bond lengths
are given in \AA.}

\end{figure}
\clearpage
\newpage
\begin{figure}[!ht]

\caption{Panels (a) and (b) show the four H$_2$ molecules adsorbed
on each Li atom in Li$_{12}$C$_{60}$-C$_5$-Li$_{12}$C$_{60}$ and
Li$_{12}$C$_{60}$-C$_8$-Li$_{12}$C$_{60}$. }
\end{figure}
\clearpage
\newpage
\begin{figure}[!ht]

\includegraphics[width=0.5\textwidth]{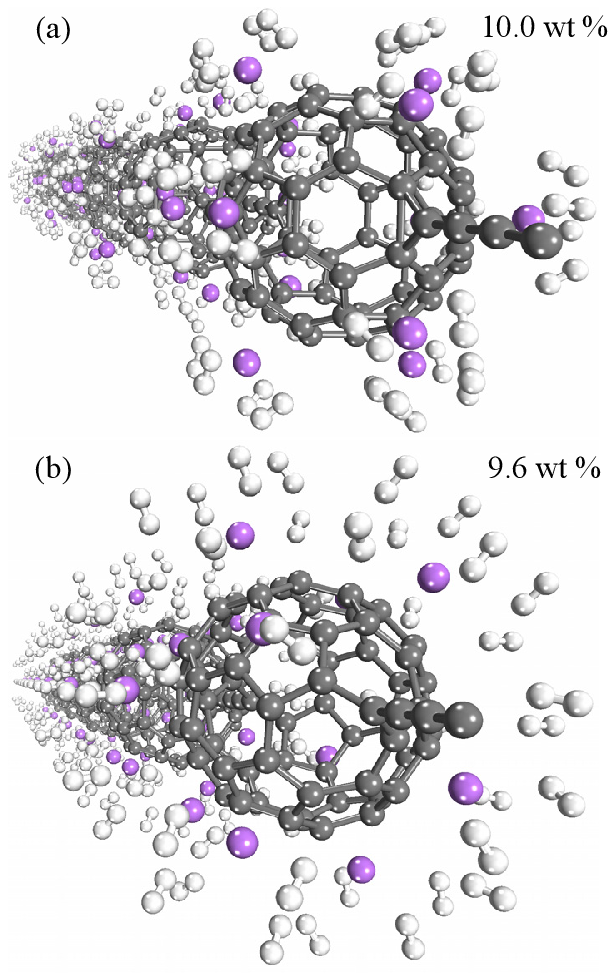}
\caption{Panels (a) and (b) show the H$_2$ molecules adsorbed on the
periodic polymers composed of
Li$_{12}$C$_{60}$-C$_5$-Li$_{12}$C$_{60}$ and
Li$_{12}$C$_{60}$-C$_8$-Li$_{12}$C$_{60}$, respectively.}

\end{figure}
\clearpage
\newpage
\end{CJK*}
\end{document}